\documentclass{llncs}
\usepackage{llncsdoc}

\usepackage{amsmath}
\usepackage{amsfonts}

\usepackage{pdfpages}
\usepackage{placeins}

\newcommand{\aref}[1]{\hyperref[#1]{Appendix~\ref{#1}}}

\usepackage{enumerate}

\newtheorem{lem}{Lemma}
\newtheorem{thm}{Theorem}
\newtheorem{cor}{Corollary}

\usepackage[stable]{footmisc}

\usepackage{tikz}
\usetikzlibrary{decorations.pathreplacing}
\usetikzlibrary{chains}
\usetikzlibrary{shapes.geometric}
\usetikzlibrary{arrows}
\usetikzlibrary{shapes}

\usepackage{float}

\usepackage{hyperref}
\usepackage{docmute}

\newcommand{\E}{\mathbb{E}}

\makeatletter
\newcommand{\IgnoreIfEmpty}[1]{%
  \begingroup
  \sbox0{#1}%
  \ifdim\wd0=\z@
    \endgroup
    \expandafter\@gobble
  \else
    \endgroup
    \expandafter\@firstofone
  \fi}
\makeatother

\newcommand{\term}[1]{\emph{#1}}

\newenvironment{cheap}{\begin{center}\begin{heap}}{\end{heap}\end{center}}

\usepackage{ifmtarg}
\makeatletter
\newcommand{\ifnonempty}[2]{\@ifmtarg{#1}{}{#2}}
\makeatother

\newcounter{phantom-fig-counter}

\newcounter{lc}
\newcounter{rc}

\newcounter{hc}

\newcounter{rcc}
\newcounter{lcc}

\tikzstyle{flowarrow}=[->,thick]
\tikzstyle{urloop}=[loop,in=45,out=20,looseness=4]

\newenvironment{heap}[1][] 
{
	\begin{tikzpicture}[anchor=center] 
	\node[draw,thick,align=left]\bgroup
	
	\ifnonempty{#1}{Heap #1 \\}
	\ifnonempty{#1}{\vspace{-0.5em}} \\
	
	\begin{tikzpicture}
	\begin{scope}[start chain=heapchain\thehc, node distance = 4mm]

    \tikzstyle{treenode}=[circle, draw, thick, edge from parent=thick]
    \tikzstyle{textnode}=[minimum width=0.9cm,inner sep=0.5mm]
        \tikzstyle{subtree}=[draw,dashed, isosceles triangle,shape border rotate=90, minimum width=8mm]
    \tikzstyle{edge from parent}=[draw,thick]
    
    \tikzstyle{level 1}=[sibling distance=7mm, level distance = 10mm]
    \tikzstyle{level 2}=[level distance = 1.1cm]
    \tikzstyle{leaf}=[fill]
    
    \tikzstyle{underbrace}=[decorate,decoration={brace,mirror,raise=2.3mm,amplitude=2.5mm},thick]

	\newcommand{\badstate}{
	\node [on chain, treenode, leaf] {};
	\node [on chain, treenode, leaf] {} child {node [treenode,leaf] {}};
	\node [on chain, treenode, leaf,xshift=2mm] {}
		child {node [treenode,leaf] {}}
		child {node [treenode,leaf] {}};
	\node [on chain,yshift=-5mm] {\dots};
	\node [on chain, treenode, leaf,xshift=2mm,yshift=5mm] {}
		child {node (desired-left-\thelcc) [treenode,leaf] {}}
		child {
			node {\dots}
			edge from parent [draw=none] {}
		}
		child {node (desired-right-\thercc) [treenode,leaf] {}};
		
	\draw [underbrace] (desired-left-\thelcc.south west) -- (desired-right-\thercc.south east)
		node [midway,below=5mm] {$ \sqrt{n}$ children};
		
    	\stepcounter{rcc}
    	\stepcounter{lcc}
	}

    \newcommand{\extratree}[3]{
    	\node [treenode,textnode,on chain=heapchain\thehc] {##1}
        	child { node (leftchild\thelc) [treenode,leaf] {} }
        	child {
        		node {\dots}
        		edge from parent [draw=none] {}
        	}
        	child { node (rightchild\therc) [treenode,leaf] {} }
        	##3
    	;
    
    	\draw [underbrace] (leftchild\thelc.south west) -- (rightchild\therc.south east)
    		node [midway,below=5mm] {##2}
    	;
    	
    	\stepcounter{rc}
    	\stepcounter{lc}
    }
    
    \newcommand{\flattree}[2]{\extratree{##1}{$##2$ children}{}}
    
    \newcommand{\onetree}[1]{
    	\customonetree{##1}{}
    }
    
    \newcommand{\customonetree}[2]{
    	\node [treenode, textnode, on chain=heapchain\thehc,##2] {##1};
    }
    
}
{
\end{scope}
\stepcounter{hc}
\end{tikzpicture}
\egroup;
\end{tikzpicture}
}

\begin{document}
\title{Replacing Mark Bits with Randomness in Fibonacci Heaps}

\author{Jerry Li \and John Peebles}
\institute{MIT \\ {\small \email{\{jerryzli, jpeebles\}@mit.edu}}}

\date{\today}
\maketitle

\begin{abstract}
A Fibonacci heap is a deterministic data structure implementing a priority queue with optimal amortized operation costs. An unfortunate aspect of Fibonacci heaps is that they must maintain a ``mark bit'' which serves only to ensure efficiency of heap operations, not correctness. Karger proposed a simple randomized variant of Fibonacci heaps in which mark bits are replaced by coin flips. This variant still has expected amortized cost $O(1)$ for insert, decrease-key, and merge. Karger conjectured that this data structure has expected amortized cost $O(\log s)$ for delete-min, where $s$ is the number of heap operations.

We give a tight analysis of Karger's randomized Fibonacci heaps, resolving Karger's conjecture. Specifically, we obtain matching upper and lower bounds of $\Theta(\log^2 s / \log \log s)$ for the runtime of delete-min. We also prove a tight lower bound of $\Omega(\sqrt{n})$ on delete-min in terms of the number of heap elements $n$. The request sequence used to prove this bound also solves an open problem of Fredman on whether cascading cuts are necessary. Finally, we give a simple additional modification to these heaps which yields a tight runtime  $O(\log^2 n / \log \log n)$ for delete-min.
\end{abstract}


\section{Introduction}

It is natural to explore the space of possible designs for common data structures. Doing so allows one to consider simpler alternative designs and gain more insight into whether particular features of a design are necessary or extraneous.

A natural class of data structures that is amenable to this sort of study is those that store additional information whose sole purpose is to ensure efficiency rather than correctness. The defining characteristic of such \emph{extraneous data} is that the data structure still functions correctly---but perhaps more slowly---if the extraneous data is corrupted.

There are numerous data structures that posses extraneous data. For example, in red-black trees \cite{guibas_dichromatic_1978}, the color of a node is extraneous data because even if we adversarially change it, the tree will still answers queries correctly---though perhaps more slowly. The balance factor of nodes in AVL trees and the mark bits of nodes in Fibonacci heaps are also extraneous data \cite{adelson-velskii_algorithm_1962,fredman_fibonacci_1987}.

In this paper, we characterize the extent to which the extraneous ``mark bit'' data contributes to the performance of Fibonacci heaps. More specifically, we give a tight analysis of what happens to asymptotic performance if one replaces the mark bits with random bits (see \autoref{sec:defs} for details).

This is interesting for three reasons. First, replacing mark bits with random bits simplifies the design of Fibonacci heaps because there is no longer a need to store any mark bits. Second, our results can also be interpreted as an analysis of the performance of Fibonacci heaps under random corruption of mark bits. Third, our results solve an open problem of Fredman on whether cascading cuts are necessary in Fibonacci heaps \cite{fredman_binomial_2005}.

\subsection{Related Work}\label{sec:related}

The randomized variant of Fibonacci heaps studied in this paper were first proposed by Karger in unpublished work in 2000 \cite{karger_manuscript}. However, Karger's analysis of the performance of these Fibonacci heaps---which we'll call \term{randomized Fibonacci heaps}---was not tight. Specifically, Karger proved an upper bound of $O(\log^2 s)$ on the expected amortized cost of delete-min where $s$ is the total number of Fibonacci heap operations performed so far. (It is easy to see that the expected amortized cost of all other operations is $O(1)$.) In terms of lower bounds, none better than the trivial sorting lower bound was known.

Following Karger's initial work, the analysis of Karger's randomized Fibonacci heaps has a somewhat amusing history. Hoping to encourage somebody to obtain a tight analysis of delete-min, Karger added this as a recurring bonus problem in MIT's annual graduate algorithms course \cite{karger_personal}. As a result, virtually every graduate student to go through MIT's theory group in the past 15 years has at least seen this problem, and many have actively worked on it.

Despite this attention, relatively little progress was made. We initially thought there had been none at all. After posting this paper, however, we were informed of two unpublished results by Eric Price that had never been posted anywhere. These consist of two bounds in terms of $s$: a ``lower bound'' weaker than ours and an upper bound that is essentially the same as ours. Price gives an adversary that queries the randomized Fibonacci heap such that it must use $\Omega (\log^2 s / \log \log s)$ expected amortized time per delete-min. However, Price's ``lower bound'' cheats by allowing the adversary's request sequence to change depending on the random choices made by the randomized Fibonacci heap. To the best of our knowledge, all algorithms that employ Fibonacci heaps don't need to inspect their private state. Thus, in all settings we are aware of, Price's lower bound does not apply. Indeed, such results are not typically described as ``lower bounds'' in the data structures literature. Moreover, Price does not give any results in terms of $n$, the number of elements in the randomized Fibonacci heap.

Shortly after this paper was posted, Kaplan, Tarjan, and Zwick posted a paper which analyzes different variants of Fibonacci heaps \cite{DBLP:journals/corr/KaplanTZ14}. Their work independently solved Fredman's open problem regarding the necessity of cascading cuts using similar techniques to ours.

More broadly, there are several data structures that have been studied which implement priority queues and achieve the same asymptotic performance as Fibonacci heaps. These include \cite{peterson_1987}, \cite{driscoll_relaxed_2013}, \cite{hoyer_1995}, \cite{takaoka_2-3_2003}, \cite{kaplan_thin_2008}, \cite{elmasry_violation_2010}, \cite{haeupler_rankpairing_2011}, \cite{chan_quake_2013}. Additionally, there are many other works that deal with pairing F-heaps and their variants; eg., \cite{fredman_pairing_1986}, \cite{pettie_pairing_2005}, \cite{elmasry_pairing_2009}. Pairing F-heaps offer slightly worse asymptotic performance than F-heaps but are often faster in practice. 



\subsection{Terminology}
We will differentiate between Fibonacci heaps as defined in \cite{fredman_fibonacci_1987} and Karger's randomized F-heaps by referring to the former as \emph{standard} Fibonacci heaps and the latter as  \emph{randomized} Fibonacci heaps. However, when the data structure we are referring to is clear from context, we may simply call it an F-heap.

For variables, $s$ will always refer to the total number of operations that have been executed on an F-heap, and $n$ will refer to the number of elements stored in the F-heap.

\subsection{Our Contributions}
We fully resolve Karger's question, giving a tight analysis of randomized F-heaps. We give a lower bound of $\Omega(\log^2 s / \log \log s)$ on the worst-case expected amortized runtime of delete-min. We also obtain a matching upper bound of $O(\log^2 s / \log \log s)$. Importantly, our lower bounds employ only non-adaptive request sequences which do not depend on the random outcomes of F-heap operations. Thus, in contrast to Price's work, our results truly are a lower bound on the amortized runtime of Karger's F-heaps.

The above two bounds are in terms of the number of F-heap operations $s$. In terms of the F-heap size $n$, we give a lower bound of $\Omega(\sqrt{n})$. (Previous work on pairing F-heaps implies a matching upper bound \cite{fredman_pairing_1986}.) The request sequence used to prove this lower bound gives an affirmative answer to the open question posed by Fredman on whether cascading cuts are necessary for performance in F-heaps.

Finally, we give a simple modification that improves the expected amortized performance to $\Theta(\log^2 n / \log \log n)$ by periodically rebuilding the F-heap.


\subsection{Roadmap}
In \autoref{sec:defs}, we review the basic properties of standard F-heaps and define randomized F-heaps. In \autoref{sec:ub-lg2n}, we prove the tight $O(\log^2 s / \log \log s)$ upper bound on the expected amortized cost of delete-min in randomized F-heaps, where $s$ is the number of F-heap operations. In \autoref{sec:lb-sqrtn}, we give a tight lower bound of $\Omega(\sqrt{n})$, where $n$ is the F-heap size. This bound serves as a warmup to our more challenging lower bound in the next section. In \autoref{sec:lb-logs}, we give a lower bound of $\Omega(\log s / \log \log s)$ on the cost of delete-min. In \autoref{sec:fix}, we give a simple modification to Karger's randomized F-heaps which improves the performance of delete-min to $O(\log^2 n / \log \log n)$, replacing the $s$ in the runtime with an $n$. We also show how to extend our work in \autoref{sec:lb-sqrtn} to yield a matching lower bound. In \autoref{sec:conclusion}, we conclude and give possible directions for future work. In \aref{app:fredman}, we explain how we resolve Fredman's open question. The rest of the appendix is proofs and large figures that could not be included in the main paper due to page limits.

\section{Background}\label{sec:defs}

A standard Fibonacci heap is a data structure that implements a priority queue and supports the operations insert, merge (or meld), decrease-key, and delete-min. The amortized runtimes of the first three operations is $O(1)$ and the amortized runtime of delete-min is $O(\log n)$
where $n$ is the F-heap size.

We will generally assume that the reader is familiar with the basic design and analysis of F-heaps. Those wishing to review this information may refer to the original paper \cite{fredman_fibonacci_1987} or any typical algorithms textbook.

Recall that each node in an F-heap allocates one bit of data called a mark bit. The only operation that uses the mark bit is the decrease-key operation. Specifically, the decrease-key operation starts by updating the key of the desired node and promoting it into the root list. Then, it starts from the node's former parent and walks up the tree, promoting nodes to the root list until it encounters a node with an unset mark bit. It then sets this node's mark bit and clears the mark bits of all nodes it promoted.

Karger defined \emph{randomized F-heaps} as follows. A randomized F-heap behaves exactly like an F-heap with one exception: how it decides to stop promoting nodes in the decrease key operation. Recall that standard F-heaps look at the mark bit to determine whether to stop walking up the tree. In contrast, randomized F-heaps flip a coin to make this decision.

Equivalently, one can think of a randomized F-heap as a simulation of a standard F-heap which intercepts queries to mark bits and returns random bits instead.

\section{An \texorpdfstring{$O(\log^2 s / \log \log s)$}{O(log{\textasciicircum}2 s / log log s)} Upper Bound}\label{sec:ub-lg2n}

In this section, we upper bound the expected amortized cost of the operations of randomized F-heaps. After obtaining this result, we were informed that unpublished work by Price that was never posted gives a similar proof of essentially this same theorem \cite{price_randomized_2009}.

\begin{thm}\label{thm:lg2n-ub}
The expected amortized costs for a randomized F-heap's operations are 
$O\left(\log s \log n /\log \log s \right) \leq O\left(\log^2 s / \log \log s \right)$
for delete-min and $O(1)$ for everything else.
\end{thm}

We use a simplified version of the potential function introduced in \cite{fredman_fibonacci_1987}: if $F$ is an F-heap, then we let $\Phi (F)$ be the number of root nodes in $F$. With this amortization, it is easy to see that insert, merge, and decrease-key all run in expected constant time. Thus it suffices to demonstrate that delete-min runs in expected time $O(\log s \log n / \log \log s)$. 

Recall the specification for delete-min: we (1) remove the minimum element from the list of roots, (2) add all of its children to the root list, then (3) perform consolidation by rank. If $k$ was the number of roots before the delete-min, $c$ the number of children of the deleted element, and $r$  was the maximum rank\footnote{Recall that the rank of a node in an F-heap is the number of children it has.} of the root node in the F-heap before performing step (3) above, then the real work performed is $O(k + c + r) = O(k + r)$ since $c \leq r$. The change in potential is $O(\log n - k)$, so with the correct scaling, the amortized cost of this operation is $O(r + \log n)$. Thus it suffices to show that $r \leq O(\log s \log n / \log \log s)$ in expectation.

We first upper-bound the probability that a node has lost many of its children since the last time it was in the root list. We say a non-root node $v$ in an F-heap is \term{missing} a child if the child was removed from $v$ and $v$ has not been in the root list since that time.

\begin{lem}\label{lem:union-bound}
Suppose we have an empty randomized F-heap and we intend to perform $s$ operations on it which will result in an F-heap of size $n$. Then the probability that every non-root node in the resulting F-heap is missing at most $k$ children is at least  $1-ns2^{-k}$.
\end{lem}

The proof is given in the appendix. As a corollary, we get the following:

\begin{cor}\label{cor:union-bound}
With probability at least $1-1/n$, no node in the F-heap described in the above lemma is missing more than $k=2\log n + \log s \leq 3 \log s$ children.
\end{cor}

For any integer $k \geq 2$, let $f_k (x) = x^k - x^{k - 1} -1$. It is not hard to see that $f_k (x)$ is increasing for $x \geq 1$, has a unique positive root $\lambda_k$, and that $\lambda_k > 1$. By following the analysis in \cite{fredman_fibonacci_1987} (in the proof of their Corollary 1), one can obtain the following result. We include the proof in the appendix for completeness. 

\begin{lem}\label{lem:global-to-children}
Suppose a tree in an F-heap with $n$ nodes has the property that no non-root node in the tree is missing more than $k$ children. Then the root has rank $O(\log_{\lambda_k} n)$.
\end{lem}

We also need a technical lemma about the behavior of $\lambda_k$, whose proof we defer to the appendix.

\begin{lem}\label{lem:lambda-k}
For $k$ sufficiently large, $1/\log \lambda_k \leq 2 k/\log k.$
\end{lem}

Now we can prove the main theorem of this section.

\begin{proof}[of \autoref{thm:lg2n-ub}]
Insert, merge, and decrease-key are obviously $O(1)$ so we focus our attention to demonstrating the bound for delete-min. The expected amortized cost of delete-min is at most the maximum rank $r$ of any root node. By \autoref{lem:global-to-children}, $r \leq O(\log_{\lambda_k} n)$ where $k$ is a bound on how many children are missing from any non-root node in the tree. We can break up $\E[r]$ into two terms and bound them separately. We have,

\[\E[r] =  \Pr [k \geq 3 \log s] \cdot \E [r | k \geq 3 \log s] + \Pr [k < 3 \log s] \cdot \E [r | k < 3 \log s]. \]
The first term is bounded by $n \mbox{Pr} [ k \geq 3 \log s] \leq 1$ by \autoref{cor:union-bound}. The second is bounded by
\[\E [r | k < 3 \log s] \leq \log n/\log \lambda_{3 \log s} = O(\log s \log n /\log \log s)\]
by \autoref{lem:lambda-k}. Thus, the total expected amortized cost of delete-min is $O(\log s \log n / \log \log s)$.
\end{proof}

\section{An \texorpdfstring{$\Omega(\sqrt{n})$}{Omega(sqrt(n))} Lower Bound}\label{sec:lb-sqrtn}

The following section is dedicated to the proof of the following lower bound:

\begin{thm}\label{thm:sqrtn-lb}
There exists a request sequence for randomized F-heaps whose expected cost is $\Omega(\sqrt{n})$ per operation on average, where $n$ is the size of the F-heap.
\end{thm}

Our proof of this result also proves that so-called ``cascading cuts'' are necessary in Fibonacci heaps, solving an open problem of Fredman. See \aref{app:fredman} for details.

It is worth clarifying what we mean when we say ``$\Omega(\sqrt{n})$ per operation on average'' since the F-heap size can change from operation to operation. Formally, this means the sum of the square roots of the F-heap sizes before each operation divided by the number of operations.

Note that the analysis used in Section 2 of \cite{fredman_pairing_1986} proves a matching upper bound. We also remark that while the expected cost of each operation in the request sequence is $\Omega(\sqrt{n})$ on average per operation, the request sequence has exponential length. We rectify this and obtain a tight bound for the expected amortized cost in terms of $s$ in Section 5.

Notice that the theorem is equivalent to saying that there is a request sequence such that---no matter how one tries to amortize the cost of the operations---there will always be an operation with cost $\Omega(\sqrt{n})$.\footnote{To see this, let the average per operation cost be $c_1$ and the maximum amortized cost of an operation be $c_2$. Then for the total cost $c_0$ of all operations in a request sequence of length $s$, we have $c_1s = t \leq c_2 s$. Thus, $c_1 \leq c_2$.}

While it is easy to slightly modify randomized F-heaps to ``get around'' this lower bound, we include this construction for three reasons. First, it applies to Karger's randomized F-heaps as they were originally formulated. Second, it is a good warm-up to the more complicated construction in \autoref{sec:ub-lg2n}, which is extended in \autoref{sec:fix} to apply even to these modified F-heaps---where $s$ is replaced with $n$ in the statement of the bound. Finally, the request sequence we construct solves Fredman's open question about the necessity of cascading cuts; see \aref{app:fredman} for more details.

The main idea is that by using a very large number of requests, we can force the F-heap into a very bad configuration with high probability. In particular, we exhibit a configuration which we call the \emph{bad state} shown in \autoref{fig:bad-state} below.

Formally, the \term{bad state of rank $\sqrt{n}$} is an F-heap with trees of rank $i$ for all $i$ from $0$ through $\sqrt{n}$ where all trees have height $1$, except the rank $0$ tree which has height $0$. For simplicity, we assume $\sqrt{n}$ is integral. Notice that the total number of F-heap elements is $\Theta(n)$. Thus, by an appropriate variable substitution, we can equivalently think of this as an $n$ element F-heap where the highest-rank node has $\Theta(\sqrt{n})$ children.

\begin{figure}[h!]
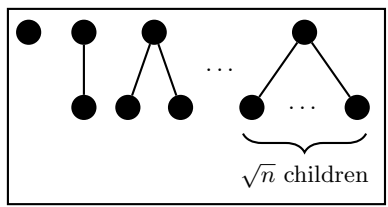

\begin{cheap}
\badstate
\end{cheap}
\caption{The bad state of rank $\sqrt{n}$. }
\label{fig:bad-state}
\end{figure}

The bad state has the following two key properties, which we encapsulate in the following two lemmas:

\begin{lem}
\label{lem:bad-state}
There exists a constant length sequence of operations which---when applied to an F-heap in the bad state of rank $\sqrt{n}$---returns the F-heap to the bad state and takes $\Omega(\sqrt{n})$ time to execute.
\end{lem}

\begin{lem}
\label{lem:slow-construction}
There exists a finite length request sequence which, when applied to an empty F-heap, results in an F-heap in the bad state of rank $\sqrt{n}$ with probability at least $1/2$.
\end{lem}



Together, these properties imply \autoref{thm:sqrtn-lb}.

\begin{proof}[of \autoref{thm:sqrtn-lb} assuming \autoref{lem:bad-state} and \autoref{lem:slow-construction}]
Fix an $n$. Construct a request sequence as follows. First, use \autoref{lem:slow-construction} to construct the first part of the request sequence. With probability at least $1/2$, this result in an F-heap in the bad state of rank $\sqrt{n}$. 
Moreover, this takes $S$ operations to execute, where $S$ is finite, known, and depends only on $n$. 
Then, follow it with a $S$ copies of the request sequence guaranteed by \autoref{lem:bad-state}. 
Conditioning on the event that the first part of the request sequence resulted in an F-heap in the bad state of rank $\sqrt{n}$, by \autoref{lem:bad-state}, each copy takes $O(\sqrt{n})$ time to execute. 
Thus we execute $O(S)$ operations on the F-heap, and with probability at least $1/2$, the operations take at least $O(S \sqrt{n})$ time. 
Therefore the expected average cost of executing this request sequence is $\Omega(\sqrt{n})$ per step on average.
\end{proof} 

Thus, all that is left is to prove \autoref{lem:bad-state} and \autoref{lem:slow-construction} which we do in the following two subsections, respectively.

\subsection{Proof of \autoref{lem:bad-state}}

From the bad state, it is straightforward to force the F-heap to spend $\Omega(\sqrt{n})$ time on a delete-min. In this subsection, we prove this fact.


\begin{proof}[of \autoref{lem:bad-state}]
Consider the following request sequence:
\begin{enumerate}
\item Add  two elements $t_1,t_2$ smaller than every element in the F-heap with $t_1<t_2$.
\item Delete-min twice.
\end{enumerate}
Applying this procedure to an F-heap in the bad state of rank $\sqrt{n}$ yields the cycle of states shown in \autoref{fig:bad-sequence} (see appendix) deterministically. Notice that the state of the F-heap after applying these operations is unchanged. Moreover, it is clear that the last delete-min operation in the procedure takes $\Theta(\sqrt{n})$ time.
\end{proof}


\subsection{Proof of \autoref{lem:slow-construction}}

This subsection proves \autoref{lem:slow-construction}.


Call a tree of height 1 where the root has $c$ children the \term{$c$-star}, so that the bad state consists of one $c$-star, for each $0 \leq c \leq  \sqrt{n}$.


We will show how to force the F-heap to construct the bad state by forcing it to construct each $c$-star in the bad state in order from large $c$ to small. Specifically, we will use the following lemma
 
\begin{lem}
\label{lem:flat-tree}
For every $c$ and $\epsilon > 0$, there exists a sequence of operations which (starting from an empty F-heap) results in an F-heap which is a $c$-star with probability at least $1-\epsilon$, and which at no point ever constructs a node with rank $> c$.
\end{lem}

We first explain why \autoref{lem:flat-tree} implies \autoref{lem:slow-construction}.

\begin{proof}[of \autoref{lem:slow-construction} assuming \autoref{lem:flat-tree}]
Our request sequence is obtained by taking the sequences obtained from \autoref{lem:flat-tree} for each $c$ from $0$ through $\sqrt{n}$ with $\epsilon$ sufficiently small, then concatenating the sequences in order from largest $c$ to smallest $c$. It is easy to see that this sequence results in the desired F-heap.
\end{proof}

\begin{proof}[of \autoref{lem:flat-tree}]
We proceed by induction on $c$. For $c=0$, simply start with an empty F-heap and insert $u$. This results in the desired F-heap with probability $1$. Inductively, suppose the statement is true for $c = k$. Fix $\epsilon > 0$. By induction, there is a request sequence which produces a $k$-star with probability at least $\sqrt{1 - \epsilon}$. Below, we describe a request sequence which constructs a $(k  +1)$-star from a $k$-star with probability at least $\sqrt{1 - \epsilon}$. Then by concatenating this request sequence to the one obtained via induction, we produce the desired request sequence. In particular, this request sequence gives rise to the desired F-heap with probability at least $(\sqrt{1 - \epsilon})(\sqrt{1 - \epsilon})=1-\epsilon$ as desired.

Assume the F-heap is a $k$-star. Now insert a node $v$ with $v>u$. This results in the F-heap shown below.

\begin{cheap}[$H$]
\flattree{$u$}{$k$}
\onetree{$v$}
\end{cheap}

Consider the following procedure which we will apply a large number of times.
\begin{enumerate}
\item Add $2^k - 1$ nodes $s_1,\ldots,s_{2^k-1}$ such that $u < v < s_1 < s_2 < \ldots < s_{2^k-1}$.
\item Add a node $t$ smaller than all other nodes in the F-heap and perform a delete-min. (This results in $t$ being removed and the rest of the nodes being consolidated.)
\item For all $1 \leq i \leq 2^k - 1$, decrease the key of $s_i$ to be minimum in the F-heap and delete-min, removing it. The order is arbitrary.
\end{enumerate}

Given an F-heap $H$ as shown in Figure 2, if we apply this procedure over and over again, the state of $H$ after any particular application of the procedure is given by the Markov process shown by the flowchart in \autoref{fig:hlv}. A more detailed step-by-step version of the flowchart is given in \autoref{fig:llv} in the Appendix.

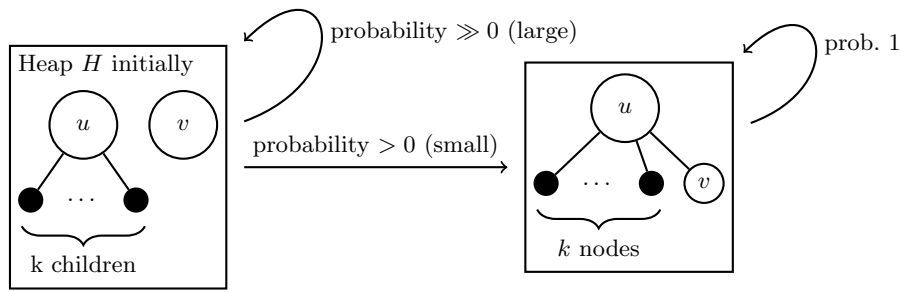
\begin{figure}[htb]
\begin{tikzpicture}
\node (hlv-startstate) {
\begin{heap}[$H$ initially]
\flattree{$u$}{$k$}
\onetree{$v$}
\end{heap}
};

\node (hlv-endstate) [right=3.5cm of hlv-startstate] {
\begin{heap}
\extratree{$u$}{$k$ nodes}{child { node [treenode] {$v$} }}
\end{heap}
};

\path (hlv-endstate)   edge[flowarrow,urloop] node [below,right]  {prob. $1$} (hlv-endstate);
\path (hlv-startstate)   edge[flowarrow,urloop] node [above,right]  {probability $\gg 0$ (large)} (hlv-startstate);
\path (hlv-startstate)   edge[flowarrow] node [above] {probability $>0$ (small)} (hlv-endstate);
\end{tikzpicture}

\caption{High-Level description of how one iteration of our procedure works. Each box represents a state of the F-heap and each arrow represents the probability of going from one state to the other after applying steps 1--3 once. After a large number of applications, we will get stuck in the state on the right with high probability.}
\label{fig:hlv}
\end{figure}


Notice that $H$ always has a positive probability of gaining a single extra child (and no extra descendants), resulting in the F-heap we are trying to create. Furthermore, once $H$ enters this state, it will never leave. Notice additionally that in none of these possible transitions do we ever produce a tree with rank greater than $k + 1$. As such, if we apply the procedure a sufficiently large number of times---and provided $H$ had the structure shown in Figure 2---we can construct a sequence of operations that gives the desired resulting $H$ with probability arbitrarily high. By repeating this request sequence sufficiently many times such that this probability is at least $\sqrt{1-\epsilon}$, we are done.
\end{proof}

\section{The \texorpdfstring{$\Omega(\log^2 s / \log \log s)$}{Omega(log{\textasciicircum}2 s / log log s)} lower bound}\label{sec:lb-logs}

This section is devoted to proving the following theorem:

\begin{thm}\label{thm:log2n-lb}
There exists a request sequence for randomized F-heaps whose expected cost is $\Omega(\log^2 s / \log \log s)$ per operation on average, where $s$ is the number of F-heap operations.
\end{thm}

Our approach to this bound has the similar structure to \autoref{thm:sqrtn-lb}: get the F-heap into a ``bad'' state then have it perform a costly operation repeatedly. However, to prove that bound, we constructed an exponentially long request sequence. The challenge in proving the present bound is that we now need a subexponential length request sequence.

\begin{figure}
\centering{
\includegraphics[scale=0.55]{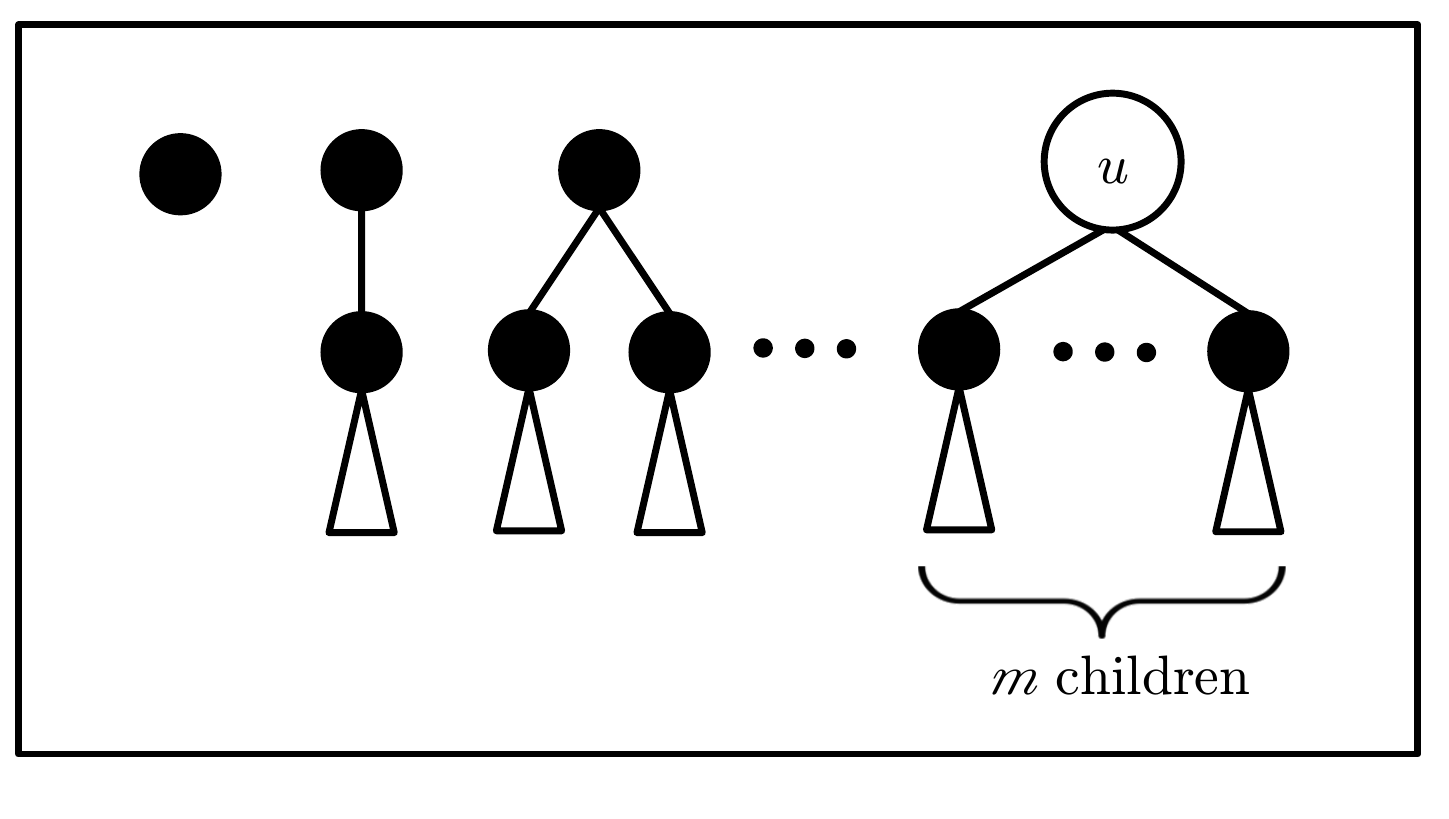}
}
\caption{A generalized bad state.}\label{fig:closed-desired}
\end{figure}

For this bound, the ``bad'' state we will force the F-heap into is defined as a \term{generalized bad state} of rank $m$ and is shown in \autoref{fig:closed-desired}.\footnote{More specifically, we will force the F-heap into a specific known state which is of the form shown in the figure.} Formally, an F-heap is in a generalized bad state of rank $m$ if it has $m + 1$ root nodes, where the $i$th root node has rank $i$, for $0 \leq i \leq m$. Once we get the F-heap into a state of this form, we will use an analog of \autoref{lem:bad-state} to make the F-heap perform costly operations, just as in the proof of \autoref{thm:sqrtn-lb}.

The analogs of \autoref{lem:bad-state} and \autoref{lem:slow-construction} we will use are the following:

\begin{lem}\label{lem:bad-state-redux}
There exists a constant length sequence of operations which---when applied to an F-heap in a generalized bad state of rank $m$---returns the F-heap to a generalized bad state of rank $m$ and takes $\Omega(m)$ time to execute.
\end{lem} 

\begin{proof}
The sequence of operations and proof is exactly the same as in \autoref{lem:bad-state}.
\end{proof}

\begin{lem}\label{lem:fast-construction}
There exists a request sequence of length $2^{O(\sqrt{m \log m})}$ which, when applied to an empty F-heap, results in an F-heap in a generalized bad state of rank $m$ with probability at least $1/2$.
\end{lem}

\begin{proof}[Proof of \autoref{thm:log2n-lb} assuming \autoref{lem:bad-state-redux} and \autoref{lem:fast-construction}]
Fix an $m$. Construct a request sequence as follows: use \autoref{lem:fast-construction} to construct the first part of the request sequence with length $\ell(m)=2^{O(\sqrt{m \log m})}$. Follow it with $\ell(m)$ copies of the constant length request sequence given by \autoref{lem:bad-state-redux}. 
This request sequence makes $\Theta( \ell (m) )$ requests and takes $\Omega (\ell (m) m)$ time, thus the average time per request is $\Omega (m)$.
Letting $s = \ell (m)$ so that $m = \ell^{-1} (s)$, we see that executing $s$ operations takes $\Omega (m) = \Omega (\ell^{-1} (s))$ time per operation on average.
Since $\ell^{-1}(s)=\Omega(\log^2 s / \log \log s)$, this completes the proof.
\end{proof}

The proof of \autoref{lem:fast-construction} is given in the appendix.

\section{Going from \texorpdfstring{$\Theta(\log^2 s / \log \log s)$}{Theta(log{\textasciicircum}2 s / log log s)} to \texorpdfstring{$\Theta(\log^2 n / \log \log n)$}{Theta(log{\textasciicircum}2 n / log log n)}}\label{sec:fix}

In this section, we eliminate the dependence on $s$ in the runtime of randomized F-heaps via a simple change. Specifically, after every operation, we rebuild the F-heap with probability $1/n$. Rebuilding is done as follows: Create a new randomized F-heap, and insert all elements from the old F-heap into the new F-heap. We refer to these self-rebuilding F-heaps as \emph{augmented} randomized F-heaps.

\begin{thm}
\label{thm:fix-ub}
The augmented randomized F-heap has worst-case expected amortized runtime $O(\log^2 n / \log \log n)$ for delete-min and $O(1)$ for everything else.
\end{thm}

\begin{thm}\label{thm:s-to-n-lb}
There exists a request sequence for augmented randomized F-heaps whose expected cost is $\Omega(\log^2 n / \log \log n)$ per operation on average, where $n$ is the number of F-heap elements.
\end{thm}

\autoref{thm:fix-ub} can be proven by using the same techniques as \autoref{thm:lg2n-ub}, so we omit it. \autoref{thm:s-to-n-lb} is proven in the Appendix.

\section{Conclusion}\label{sec:conclusion}

This work gave the first tight analysis of randomized F-heaps, resolving a 15 year old question of Karger and a 10 year old open problem of Fredman. We showed that replacing the extraneous mark bit data in F-heaps hurts performance, but only by roughly a $\log$ factor. A natural question for further work is whether replacing extraneous data with randomness in other data structures like red-black trees and AVL trees also hurts their performance.

\section{Acknowledgments}

We would like to thank David Karger for making us aware of this problem and for pointing out that our analysis in \autoref{sec:ub-lg2n} actually gave us something tighter than we originally thought.

\bibliographystyle{alpha}
\bibliography{fib-heaps-paper}

\appendix

\section{Solving Fredman's Open Problem}\label{app:fredman}

Consider the folowing variant of standard F-heaps. Unlike standard F-heaps, this variant only promotes the starting node to the root list when performing decrease-key operation. (Standard F-heaps ``cascade'' up the tree, stopping when they reach an unset mark bit; see \autoref{sec:defs} for details.) Fredman posed the question of whether this variant achieves the same asymptotic performance as standard F-heaps \cite{fredman_binomial_2005}.

It is not hard to see that our request sequence and analysis in \autoref{sec:lb-sqrtn} also prove the same $\Omega(\sqrt{n})$ lower bound for this F-heap variant. Specifically, our analysis only requires that the probability of promoting the starting node's ancestors to the root list be $\leq$ $1/2$. If its ancestors are never promoted---as is the case here---all the bounds in our analysis only get better.

It is also not hard to see that if we are only concerned with proving a lower bound on this F-heap variant, the request sequence can be simplified so that it has polynomial---rather than exponential---length because we no longer need to repeatedly perform operations until we get lucky.

\section{Omitted Proofs}

\subsection{Proof of \autoref{lem:union-bound}}

\begin{proof}
We can think of the request sequence as being generated by an adversary who wishes to make many non-root nodes that are each missing at least $k$ of their children. In order to do this, the adversary will inevitably be required to cascade decrease-key operations up to the children it wants to remove. However, any decrease-key operation that gets rid of a child also has probability $1/2$ of sending the parent to the root list as well. However, the adversary gets many tries to remove $k$ children from a node, potentially up to one per delete-min operation. The probability of any try succeeding is $2^{-k}$ and the adversary gets at most $s$ tries on each of the $n$ nodes. A union bound over all tries on all nodes gives a probability of success for the adversary of no more than $ns2^{-k}$.

More formally, for any node $v$ in the F-heap, let $p_{v,t}$ denote the probability that $v$ just became a non-root node during operation $t$, \quad \footnote{IE., $v$ is a non-root node after operation $t$ and either (1) there was no operation before operation $t$ or (2)$v$ was a root node after operation $t-1$.} never returns to the root list after operation $t$, and is eventually missing at least $k$ of its children after operation $t$.

A necessary condition for this event is that $k$ of $v$'s children get promoted to the root after time $t$, but the cascade does not continue on to $v$. The probability of this is at most $2^{-k}$. So, $p_{v,t} \leq 2^{-k}$.

Then the probability $p_v$ that node $v$ is missing at least $k$ children after $s$ operations is $\sum_{i=1}^s p_{v,i} \leq s2^{-k}$.  Taking a union bound over all nodes gives an upper bound of adversary success of $ns2^{-k}$.
\end{proof}

\subsection{Proof of \autoref{lem:global-to-children}}

\begin{proof}
We will show that if no node is missing than $k$ children, every node in the F-heap of rank $r$ has at least $(\lambda_k)^r$ descendants. 
In particular, if the root node has rank $r$, we have $n \geq (\lambda_k)^r$, so $\log_{\lambda_k} n \geq r$.

Fix a node $n$ in the F-heap with $r > k$ children.  Order its children $n_1, \ldots, n_r$ in the order they were added. Let $F_i$ be the number of children $n_i$ has. For all $i$ it must have been that $n_i$ had at least $i - 1$ children when was added because $n$ had at least $i - 1$ children when $n_i$ was added; namely, $n_j$ for $j < i$. 
Since at most $k - 1$ children can be lost before the node itself is cut, this implies that $F_i \geq i - k + 1$ for $i \geq k - 1$. Notice that in the worst case this analysis is tight; if $n$ never had children other than $n_0, \ldots, n_r$ and each child lost as many children as possible without cascading, then node $n_i$ has exactly $0$ children for $i < k$ and $i - k + 1$ children for $i \geq k$.

Let $S_{r}$ be the minimum number of descendants of a node (including itself) with $h$ children. By the above, we have $S_{r} \geq \sum_{i = 0}^{k - r} S_{i} + k$ for $r \geq k$ and by definition $S_{r} \geq r + 1$ for $r < k$; moreover, in the worst case, these are equalities. Thus we have $S_r \geq S_{r - 1} + S_{r - k}$ and $S_k \geq k$. 

Setting these inequalities in this recurrence to be equalities, we find that the characteristic polynomial is $f_k (x)$. Thus, we have $S_r = (\lambda_k)^r$, as claimed.
\end{proof}

\subsection{Proof of \autoref{lem:lambda-k}}
\begin{proof}
The lemma is equivalent to the claim that for $k$ sufficiently large, $\lambda_k \geq k^{1/(2k)}$. 
Both sides of this inequality are greater than 1. 
Thus, it suffices to show that for sufficiently large $k$, we have $f_k (k^{1/(2k)}) < 0$, as then by the monotonicity of $f_k$ on $[1, \infty)$ we have that the unique zero crossing of $f_k$ in $[1, \infty)$, namely $\lambda_k$, occurs after $k^{1/(2k)}$ and thus $\lambda_k \geq k^{1/(2k)}$, as claimed.
Since $f_k (k^{1/(2k)}) = k^{1/2} - k^{(k - 1)/(2k)} - 1$ and $k^{1/2} - k^{(k - 1)/ (2k)} \to 0$ as $k \to \infty$, we arrive at the desired conclusion.
\end{proof}

\subsection{Proof of \autoref{lem:fast-construction}}

We construct our request sequence by stringing together shorter request sequences which we will call \term{subroutines}. We will be very careful to control the exact internal state of the F-heap throughout this process. To this end, we will construct our subroutines so that they have a completely \term{predictable effect} on the F-heap, even though these subroutines may invoke F-heap operations that employ randomization. In other words, we require that if we know the exact starting state of the F-heap, apply a subroutine to the F-heap, and the subroutine succeeds, we should be able to infer the exact ending state of the F-heap. From this point forward, when we say ``subroutine,'' we mean ``subroutine with predictable effect.''

Say a subroutine is an \term{evil subroutine for rank $m$} with failure probability $p_\text{evil}$ if, when given an F-heap where every root node has rank $>m$, the subroutine  produces an F-heap which is the union of the F-heap it was given and the generalized bad state of rank $m$, and it fails with probability at most $p_\text{evil}$. Our strategy is to construct an evil subroutine by making multiple calls to what we call a \term{shifting subroutine} which we define below. Our strategy for constructing a shifting subroutine will be to make multiple calls to evil subroutines, resulting in mutual recursion.

Intuitively, a shifting subroutine of rank $k$ takes an F-heap in a generalized bad state of rank $(k-1)$, looks at the node of rank $(k-1)$ in the root list, and increments its rank. Unfortunately, this subroutine requires a rather long formal definition as there are several technical conditions it must satisfy.

More formally, a \emph{shifting subroutine for rank $k$} with failure probability $p_\text{shift}$ will give the result F-heap specified below with probability at least $(1-p_\text{shift})$, assuming the preconditions given below are satisfied.

The preconditions are that the F-heap have precisely the following nodes in its root list:
\begin{enumerate}
\item exactly one node of rank $i$ in the root list for all $0 \leq i \leq (k-1)$,
\item  no nodes of rank $k$, and
\item possibly some nodes of rank $(k+1)$ or greater.
\end{enumerate}

Let $S$ be the set of trees whose roots have rank $(k+1)$ or greater prior to the shifting subroutine being applied. The result of the shifting subroutine is an F-heap which satisfies three conditions. First, it has precisely the following nodes in its root list: 
\begin{enumerate}
\item exactly one node of rank $i$ in the root list for all $0 \leq i \leq (k-2)$,
\item  no nodes of rank $k-1$,
\item exactly one node of rank $k$, and
\item possibly nodes of rank $(k+1)$ or greater.
\end{enumerate}
Second, the trees from $S$ have not been modified in any way by the shifting subroutine. Third, there are no trees whose root nodes have rank $(k+1)$ or greater besides those in $S$.

Note that this formal definition does not require the nodes in the root list of rank $<k$ to be the same before and after the subroutine is applied.

We now prove several lemmas which are necessary to prove \autoref{lem:fast-construction}.

\begin{lem}\label{lem:shifting-to-evil}
Suppose for all $k \leq m$ and any probability $p_\text{shift}$ that there exists a shifting subroutine for rank $k$ with failure probability at most $p_\text{shift}$ and length $f(m,p_\text{shift})$. Then for any probability $p_\text{evil}$, there exists an evil subroutine for rank $m$ with failure probability at most $p_\text{evil}$ and length at most $m^2 f(m,p_\text{evil}/m^2)$.
\end{lem}

\begin{proof}
Insert a lone node into the F-heap. Then, for each $i$ from $1$ through $(m-1)$ in descending order, do the following two things:
\begin{enumerate}
\item Iterate over all $j$ from $i$ through $1$ in order and append a shifting subroutine for rank $j$ and failure probability $p_\text{evil}/m^2$.
\item Insert a node into the F-heap.
\end{enumerate}
It is easy to see that the resulting sequence has the desired length and failure probability.
\end{proof}

We have established that given a shifting subroutine, we can construct an evil subroutine. We now give a construction in the reverse direction.

\begin{lem}
\label{lem:evil-to-shifting}
Let $m$ be sufficiently large. Suppose for all $k < m$ and any probability $p_\text{evil}$ that there exists an evil subroutine for rank $k$ with failure probability at most $p_\text{evil}$ and length $g(k,p_\text{evil})$. Then for any probability $p_\text{shift}$ and any $w$, there exists a shifting subroutine for rank $m$ with failure probability at most $p_\text{shift}$ and length at most $2^w \cdot \ln \frac{2}{p_\text{shift}} + g(m-w - 1,p_\text{shift}/2)$.
\end{lem}

\begin{proof}
Assume the F-heap satisfies the preconditions for a shifting subroutine of rank $m$. Let $u$ be the element in the F-heap with $m$ children. To start off, decrease-key on $u$ so that it is the smallest element in the F-heap. Then add an element $t_2$ such that $u<t_2<\text{everything else}$.

We perform the following procedure $y=2^w \cdot \ln \frac{2}{p_\text{shift}}$ times:

\begin{enumerate}
\item Insert an element $t_1$ such that $t_1<u<t_2<\text{everything else}$. Then perform a delete-min operation, which removes $t_1$ and forces a consolidation.
\item Perform decrease-key on the children of $t_2$ with ranks $m - w - 1, \ldots, m - 2$. Notice that these are exactly the elements that were previously in the root list with the same ranks.
\end{enumerate}

The results of this operation are shown in Figure \ref{fig:big_fig}. Notably, from the starting state, a single iteration of this procedure has one of two effects: with probability $1 - 2^{-w}$, it enters into a state $S_1$ where $t_2$ is made a root in one of the decrease-key operations performed in step (2), and with probability $2^{-w}$, it enters a state $S_2$ where $t_2$ remains a child of $u$, which now has $i + 1$ children. By inspection, after each application of this procedure, we always end up in state $S_1$ or $S_2$. Moreover, any iteration that starts in $S_2$ ends in $S_2$ with probability $1$, and any iteration that starts in $S_1$ ends up in $S_2$ with probability at least $2^{-w}$, and otherwise ends up back in $S_1$. Thus, if we repeat this procedure $y$ times, the probability we do not end up at state $S_2$ is at most $(1 - 2^{-w})^y \leq p_\text{shift} / 2$.

Moreover, in $S_2$, as seen in the figure, in addition to $u$, the root list exactly consists of one element with $j$ children, for each $m - w - 1 \leq j \leq m$. Thus, after exiting the loop, applying an evil subroutine for rank $m - w - 1$ with failure probability $p_\text{shift} / 2$ yields a shifting subroutine with the desired runtime and failure probability.
\end{proof}

We now prove the following lemma which immediately implies \autoref{lem:fast-construction}.

\begin{lem}\label{lem:evil-amplification}
For sufficiently large $m$, all $1 \leq j \leq m$, and any probability $p_\text{evil}$, there exists an evil subroutine for rank $m$ with failure probability $p_\text{evil}$ and length at most $\ell(m)=g(m,p_{\text{evil}})=m^{3j} 2^{m /j} \ln \frac{1}{p_\text{evil}}$. Setting $j=\sqrt{m / \log m}$ and $p_\text{evil}=1/2$ gives $\ell(m)=2^{O(\sqrt{m \log m})}$.
\end{lem}

\begin{proof}
We proceed by induction on $j$.

Base Case: For $j=1$, a sequence of length $O(2^m)$ suffices. If we add $2^{m} + 1$ nodes smaller than any other nodes in the F-heap and delete-min, we get an F-heap in a generalized bad state. 

However, if we only did this, we wouldn't be able to keep track of where all the nodes go, so the request sequence wouldn't have \emph{predictable effect}.
To get around this technical difficulty, when inserting nodes, we stop the insertion process after every insertion of a node to force a consolidation by inserting a dummy and doing a delete-min operation. 
This does not change the asymptotic behavior of the runtime or the number of operations.
By doing this, there will be at most one pair of nodes that can consolidate at each step, so the request sequence is predictable.

Induction Hypothesis: Assume the result holds for some $i = j$. Then for any probability $p_\text{evil}$, there exists an evil subroutine for rank $j$ with failure probability $p_\text{evil}$ and length at most $\ell(m)=g(m,p_\text{evil})=m^{3j} 2^{m/j} \ln \frac{1}{p_\text{evil}}$.  Now consider $i=j+1$.

By our induction hypothesis and \autoref{lem:evil-to-shifting} with $w=m/(j+1)$, we have that for any probability $p_\text{shift}$, there exists a shifting subroutine for rank $m$ with failure probability at most $p_\text{shift}$ and length at most

\[
2^{m/(j+1)} \cdot \ln \frac{2}{p_\text{shift}} + m^{3j} 2^{m\cdot \frac{j}{j+1} / j} \ln \frac{2}{p_\text{shift}} \leq 2 m^{3j} 2^{m/(j+1)} \ln \frac{2}{p_\text{shift}}.
\]

By this and \autoref{lem:shifting-to-evil}, there exists evil subroutine for rank $m$ with failure probability $p_\text{evil}$ and length at most

\[
m^2 2 m^{3j} 2^{m/(j+1)} \ln \frac{2m^2}{p_\text{evil}} \leq m^{3(j+1)} 2^{m/(j+1)} \ln \frac{1}{p_\text{evil}}
\]

for sufficiently large $m$.
\end{proof}

\subsection{Proof of \autoref{thm:s-to-n-lb}}

Notice that the only thing that prevents our request sequence given in \autoref{thm:log2n-lb} from directly applying to augmented randomized F-heaps is that the augmented F-heaps periodically rebuild themselves, messing up the F-heap state. Our strategy is to simply prevent the F-heap from rebuilding itself. Specifically, we add a very large number of nodes to the F-heap so the rebuild probability will be low.

\begin{proof}
We obtain the desired request sequence by modifying the sequence constructed in the proof of \autoref{thm:log2n-lb}. Specifically, our new request sequence is as follows: insert $\left(2^{\lceil 100 \cdot \sqrt{m \log n} \rceil} + 1\right)$ entries into the F-heap, then perform a delete-min. Provided that this final delete-min does not trigger a rebuild, the result of these operations---regardless of any rebuilding that occurs prior to the delete-min operation---is an F-heap with a single node in the root list which has $\Theta(2^{\lceil 100 \cdot \sqrt{m \log n} \rceil})$ children.

After performing the requests described above, we simply perform the requests given by \autoref{thm:log2n-lb} as usual. Recall that the large number of nodes we added prior to performing the requests from \autoref{thm:log2n-lb} are all consolidated under a single node of large rank. Notice that the rank of this node is actually larger than any possible rank of any other node in the F-heap for the entirety of the request sequence. Thus, the presence of these extra nodes will not affect any operations from \autoref{thm:log2n-lb}.

So, provided no rebuild happens after the delete-min mentioned above and no rebuild happens during any subsequent operations, this whole request sequence has expected cost $\Theta(2^{c \sqrt{m \log m}})$ for some constant $c$. Furthermore, the probability of such a rebuild happening is so small that the overall expected cost is still $\Theta(2^{d \sqrt{m \log m}})$ for some constant $d$. The desired result immediately follows.
\end{proof}

\section{Large Figures}\label{app:figs}

Each subsection in \autoref{app:figs} contains figures originally from the same-numbered section in the main paper.

\setcounter{subsection}{3}
\newcommand{\appsec}{\FloatBarrier\subsection{}}

\appsec

\begin{figure}[!h]
\begin{tikzpicture}[every node/.style={scale=0.9}]
\node (bad-seq-start) {
\begin{heap}[initial state]
\badstate
\end{heap}
};

\node (bad-seq-mid) [right=1.8cm of bad-seq-start] {
\begin{heap}
\badstate
\node [on chain] {}; 
\onetree{$t_1$}
\onetree{$t_2$}
\end{heap}
};

\path (bad-seq-start)   edge[flowarrow] node [above] {insert $t_1$, $t_2$} (bad-seq-mid);

\node (bad-seq-end) [below=8mm of bad-seq-mid]  {
\begin{heap}
\node  [on chain=heapchain\thehc,treenode] {$t_2$}
	child {node [treenode, leaf] {}}
	child {node [treenode, leaf] {} child {node [treenode,leaf] {}}}
	child {node [treenode, leaf,xshift=2mm] {}
		child {node [treenode,leaf] {}}
		child {node [treenode,leaf] {}}
	}
	child {node [xshift=3.5mm,yshift=-5mm] {\dots} edge from parent [draw=none]}
	child {node [treenode, leaf,xshift=6mm] {}
		child {node (desired-left-part) [treenode,leaf] {}}
		child {
			node {\dots}
			edge from parent [draw=none] {}
		}
		child {node (desired-right-part) [treenode,leaf] {}}
	}
;

\draw [underbrace] (desired-left-part.south west) -- (desired-right-part.south east)
	node [midway,below=5mm] {$ \sqrt{n} \;$ children};

\end{heap}
};

\path (bad-seq-mid)   edge[flowarrow] node [right] {first delete-min} (bad-seq-end);
\path (bad-seq-end)   edge[flowarrow,out=180,in=-50] node [below left] {second delete-min} (bad-seq-start);
\end{tikzpicture}
\caption{Creating many expensive delete-min operations.}\label{fig:bad-sequence}
\end{figure}
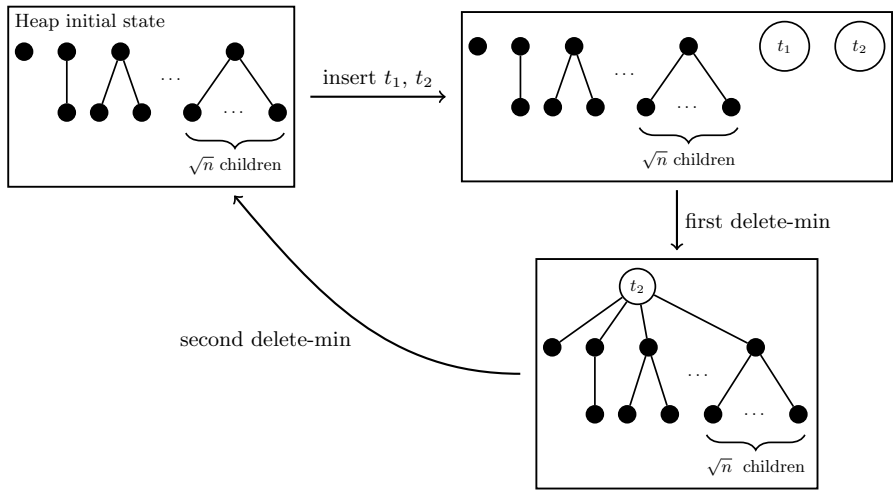

\begin{figure}
\begin{tikzpicture}[every node/.style={scale=0.9}]

\node (llv-top-start) {
\begin{heap}[$H$ initially]
\flattree{$u$}{$k$}
\onetree{$v$}
\end{heap}
};

\node (llv-top-mid) [right=1.5cm of llv-top-start] {
\begin{heap}
\flattree{$u$}{$k$}
\onetree{$v$}

\customonetree{$s_{1}$}{xshift=1mm}
\node [on chain,xshift=-4mm] {$\cdots$};
\customonetree{$s_{2^k -1}$}{scale=0.9,xshift=-4mm}
\end{heap}
};

\path (llv-top-start)   edge[flowarrow] node [above] {Step $1$} (llv-top-mid);

\node (llv-top-end) [below = 8mm of llv-top-mid] {
\begin{heap}
\extratree{$u$}{$k$ nodes}{
child { node [treenode] {$v$}
	child {node (llv-top-mid-subtree) [subtree] {} }
	}
};

\node (llv-top-mid-subtree-label) [above right=of llv-top-mid-subtree] {the $s_i$ nodes};

\path (llv-top-mid-subtree-label) edge[->,shorten >=-2.5mm,out=-45,in=20] (llv-top-mid-subtree);
\end{heap}
};

\path (llv-top-end)   edge[flowarrow,out=180,in=-30] node [below left,yshift=2mm] {Step $3$ with prob. $\gg 0$ (large)} (llv-top-start);
\path (llv-top-mid)   edge[flowarrow] node [right] {Step $2$} (llv-top-end);

\node (llv-bottom-start) [below left=of llv-top-end,yshift=7mm,xshift=-9mm] {
\begin{heap}
\extratree{$u$}{$k$ nodes}{child { node [treenode] {$v$} }}
\end{heap}
};

\path (llv-top-end)   edge[flowarrow,out=180,in=30] node [above left,yshift=-1mm] {Step $3$ with prob. $>0$ (small)} (llv-bottom-start);

\node (llv-bottom-mid) [right=1.9cm of llv-bottom-start] {
\begin{heap}
\extratree{$u$}{$k$ nodes}{child { node [treenode] {$v$} }}

\customonetree{$s_{1}$}{xshift=1mm}
\node [on chain,xshift=-4mm] {$\cdots$};
\customonetree{$s_{2^k -1}$}{scale=0.9,xshift=-4mm}
\end{heap}
};

\path (llv-bottom-start)   edge[flowarrow] node [above] {Step $1$} (llv-bottom-mid);

\node (llv-bottom-end) [below=8mm of llv-bottom-mid] {
\begin{heap}
\extratree{$u$}{$k$ nodes}{child { node [treenode] {$v$} }}

%

\node (llv-bottom-end-l1) [on chain=heapchain\thehc,xshift=2cm,subtree,anchor=north,minimum height=1.4cm,thick] {};
\node (llv-bottom-end-l2) [on chain=heapchain\thehc,subtree,xshift=2mm, minimum height=1.4cm,thick] {};
\node [on chain] {$\cdots$};
\node (llv-bottom-end-lend) [on chain=heapchain\thehc,subtree,xshift=-1mm, minimum height=1.4cm,thick] {};

\draw [underbrace] (llv-bottom-end-l1.south west) -- (llv-bottom-end-lend.south east)
	node [midway,below=5mm] {\shortstack{trees formed by the $s_i$ nodes:\\ the left tree's root has $k-1$ children,\\ the next tree's root has $k-2$ children,\\ \vdots \\ the last tree's root has $0$ children}}
;
\end{heap}
};

\path (llv-bottom-mid)   edge[flowarrow] node [right] {Step $2$} (llv-bottom-end);
\path (llv-bottom-end)   edge[flowarrow,out=170, in=-90] node [below left] {Step $3$} (llv-bottom-start);
\end{tikzpicture}

\caption{Detailed description of how one iteration of our procedure works. If only one arrow is shown, the probability of going from the state it starts at to the state it ends at is $1$. Triangles denote trees/subtrees.}
\label{fig:llv}
\end{figure}
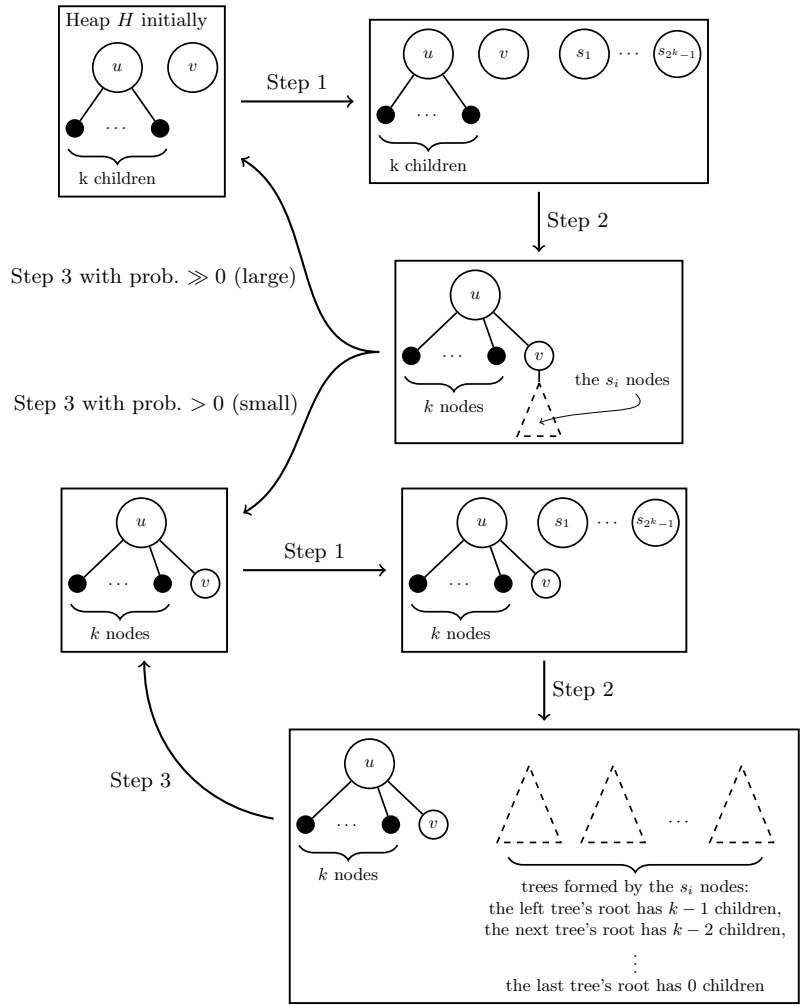

\appsec

\begin{figure}[!h]
\includegraphics[scale=0.65]{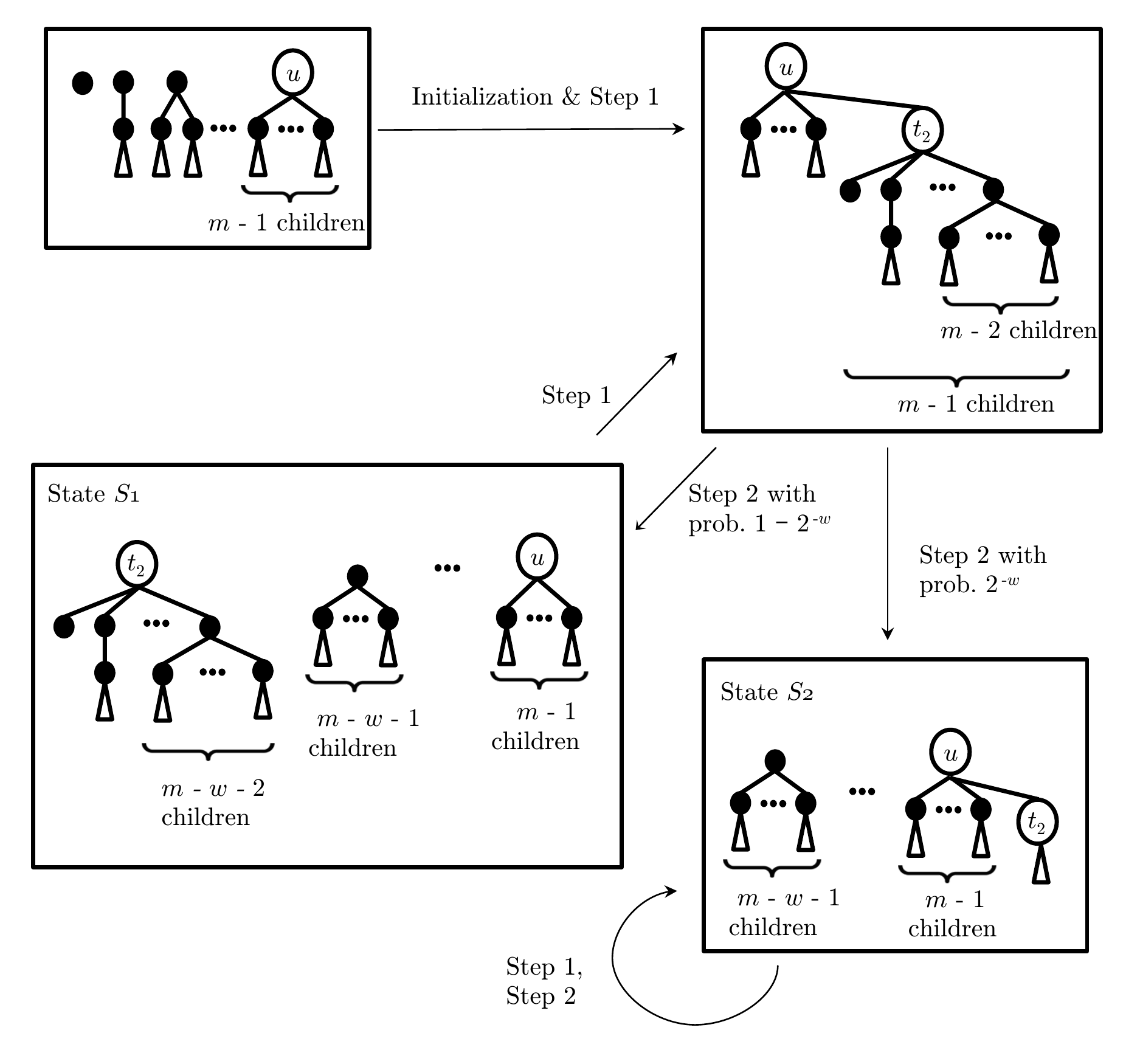}
\caption{Producing shifting subroutines via evil subroutines}
\label{fig:big_fig}
\end{figure}

\end{document}